# A Low-mass GEM Detector with Radial Zigzag Readout Strips for Forward Tracking at the EIC

Marcus Hohlmann, Matthew Bomberger, Stefano Colafranceschi, Francisco Jimenez, Mehdi Rahmani, Aiwu Zhang

*Abstract*–We present design and construction of a large low-mass Triple-GEM detector prototype for forward tracking at a future Electron-Ion Collider. In this environment, multiple scattering of forward and backward tracks must be minimized so that electron tracks can be cleanly matched to calorimeter clusters and so that hadron tracks can efficiently seed RICH ring reconstruction for particle identification. Consequently, the material budget for the forward tracking detectors is critical. The construction of the detector builds on the mechanical foil stretching and assembly technique pioneered by CMS for the muon endcap GEM upgrade. As an innovation, this detector implements drift and readout electrodes on thin large foils instead of on PCBs. These foils get stretched mechanically together with three GEM foils in a single stack. This reduces the radiation length of the total detector material in the active area by a factor seven from over 4% to below 0.6%. It also aims at improving the uniformity of drift and induction gap sizes across the detector and consequently signal response uniformity. Thin outer frames custom-made from carbon-fiber composite material take up the tension from the stretched foil stack and provide detector rigidity while keeping the detector mass low. The gas volume is closed with thin aluminized polyimide foils. The trapezoidal detector covers an azimuthal angle of 30.1 degrees and a radius from 8 cm to 90 cm. It is read out with radial zigzag strips with pitches of 1.37 mrad at the outer radius and 4.14 mrad at the inner radius that reduce the number of required electronics channels and associated cost while maintaining good spatial resolution. All front-end readout electronics is located away from the active area at the outer radius of the trapezoid. Scans of small readout boards with the same type of zigzag strip structure using highly collimated X-rays show spatial resolutions of 60-90 microns.

## I. Introduction

A future Electron-Ion Collider (EIC) will be a powerful tool to explore the properties of nuclear matter in the gluon-dominated regime via tomography of partons in nuclei. It will also measure orbital angular momenta of partons to explain the nucleon spin [1]. Current proposals of EICs in the United States are eRHIC at BNL [2] and JLEIC at Jefferson Lab [3]. The three EIC detectors currently being proposed (BeAST, ePHENIX, JLAB IP1 central detector) feature virtually identical designs for the forward and backward tracking regions. In all three detector designs, both regions are instrumented with tracker disks made from large-area GEM detector modules that provide acceptance from close to the beam pipe out to a radius of about one meter (Fig. 1). The forward tracking group of the eRD6 consortium is conducting R&D that specifically targets the development of affordable low-mass GEM detectors for this common subdetector system.

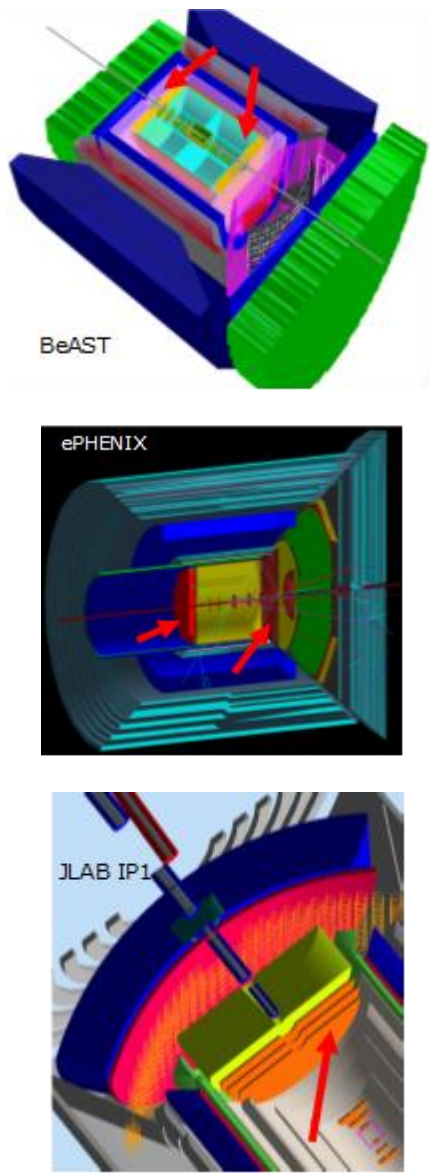

Fig. 1. Commonality of the designs for the forward and backward tracking regions with GEM detector disks (red arrows) in currently proposed EIC detectors.

Manuscript received Nov 14, 2017. This work was supported by the EIC R&D program at Brookhaven National Laboratory via the eRD6 consortium.
  M. Hohlmann, M. Bomberger, S. Colafranceschi, F. I. Jimenez, and M. Rahmani are with Dept. of Physics and Space Sciences, Florida Institute of Technology, Melbourne, FL 32901 USA (telephone 321-674-7275, e-mail: hohlmann@fit.edu).
  A. Zhang, was with Dept. of Physics and Space Sciences, Florida Institute of Technology, Melbourne, FL 32901 USA. He is now with Brookhaven National Laboratory, Upton, NY 11973 USA.

## II. DESIGN AND CONSTRUCTION OF A LOW-MASS GEM DETECTOR WITH RADIAL ZIGZAG READOUT STRIPS

In an EIC experiment, multiple scattering of tracks must be minimized so that electron tracks can be cleanly matched to calorimeter clusters and so that hadron tracks can efficiently seed RICH ring reconstruction for particle identification. Consequently, the material budget available for the forward and backward tracking detectors is a critical parameter. We have designed and constructed a large low-mass GEM detector as an option for tracking in the forward/backward EIC detector regions. It is read out with radial zigzag strips to minimize the number of required electronics channels and associated cost while maintaining good spatial resolution

As an innovation, this EIC GEM prototype detector implements drift and readout electrodes on thin large foils instead of on PCBs. Table 1 shows that this reduces the radiation length of the total detector material in the active area by a factor seven from typically over 4% for a PCB-based detector, e.g. the CMS GEM detector for the CMS forward muon upgrade [4], to just below 0.6% in the active area of this all-foil detector. The dominant material contribution (83%) is from the copper cladding of the five foils.

TABLE I. MATERIAL ACCOUNTING FOR THE DETECTOR PROTOTYPE COMPARED WITH A REFERENCE DETECTOR CONSTRUCTED FROM PCBS

| Detector with PCBs (e.g. CMS) | Thickness (mm) | % of Rad. Length |
|---|---|---|
| 2 PCBs (drift and readout) | 3.180 | 3.914 |
| 3 GEMs | 0.180 | 0.261 |
| Polyimide | 0.150 | 0.051 |
| Copper | 0.030 | 0.210 |
| **Total** | | **4.175%** |

| Detector with foils only (EIC) | Thickness (mm) | % of Rad. Length |
|---|---|---|
| 2 Al-Polyimide foils (gas seal) | 0.051 | 0.0184 |
| Polyimide | 0.051 | 0.018 |
| Al | 0.0002 | 0.0004 |
| 3 GEMs | 0.180 | 0.261 |
| Polyimide | 0.150 | 0.051 |
| Copper | 0.030 | 0.210 |
| 1 GEM as drift foil | 0.060 | 0.087 |
| Polyimide | 0.050 | 0.017 |
| Copper | 0.010 | 0.070 |
| Readout foil | 0.080 | 0.227 |
| Polyimide | 0.050 | 0.017 |
| Copper (15 μm each side b/c of vias) | 0.030 | 0.210 |
| **Total** | | **0.593%** |

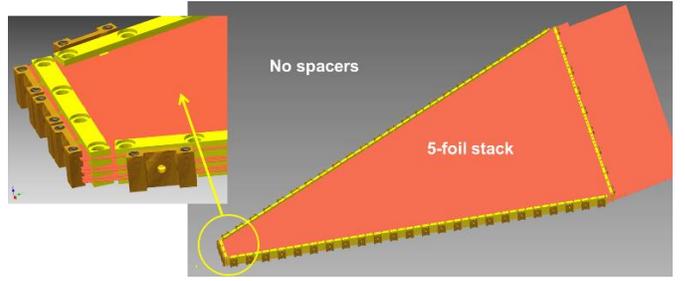

Fig. 2. Stack design with drift foil, three GEM foils, and readout foil.

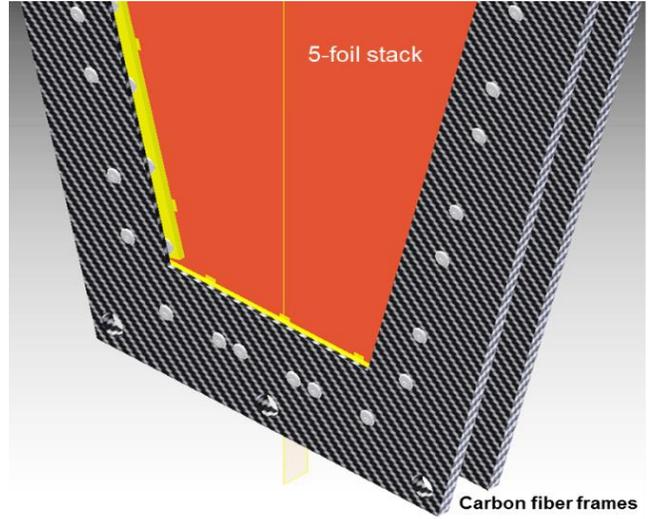

Fig. 3. Design of carbon fiber frame.

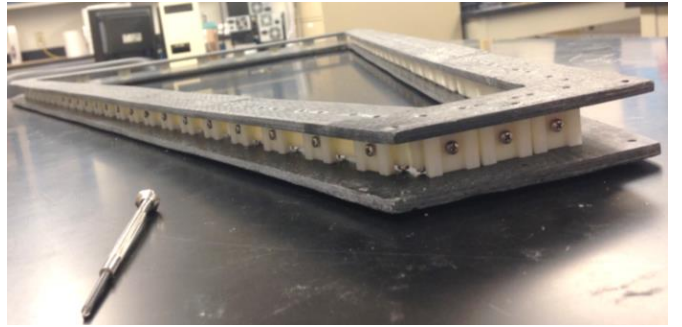

Fig. 4. Test assembly of carbon fiber prototype frame.

The drift and readout foils get stretched mechanically together with three GEM foils in a single stack (Fig. 2). This joint assembly also aims at improving the uniformity of gap sizes across the detector, which are nominally drift/transfer-1/transfer-2/induction = 3/1/2/1 mm, and consequently signal response uniformity. Thin outer frames built in-house from carbon-fiber composite material (Figs. 3,4) take up the tension from the stretched stack and provide rigidity to the detector while keeping the detector mass low. The gas volume is closed with aluminized polyimide foils. The trapezoidal detector has an active area with an azimuthal angle of 30.1 degrees and a radius that ranges from 8 cm to 90 cm.

The readout foil for the GEM detector features regular straight strips in the radially innermost sector and radial zigzag strips in the other four radial sectors (Figs. 5-7). The strip pitch in the two innermost sectors is 4.14 mrad and 1.37 mrad in the outer three sectors. The radial distance between two tips of the same zigzag strip is 0.5 mm (Fig. 5).

## III. SPATIAL RESOLUTION ACHIEVED WITH ZIGZAG READOUT STRUCTURE

The resolution obtainable with such zigzag-strip structures is measured with a highly collimated X-ray source impinging on small Triple-GEM detectors equipped with 10 cm × 10 cm test boards that feature radial zigzag strips with close-to-optimal interleaving of adjacent strips on PCBs and on a foil [5]. We observe robust charge sharing among adjacent strips, which in turn results in a fully linear spatial response. On a linear scale, strip pitches are around 1 mm and the zigzag readouts achieve linear spatial resolutions of 50-90 µm at medium gas gains around 3000 [5].

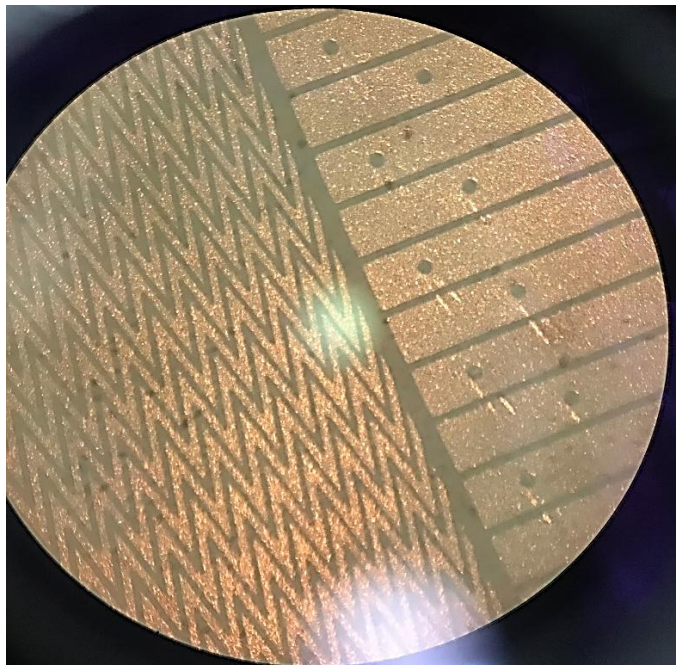

Fig. 5. Microscopic view of interface between sector 1 with straight strips (right) and sector 2 with zigzag strips (left) on the readout foil.

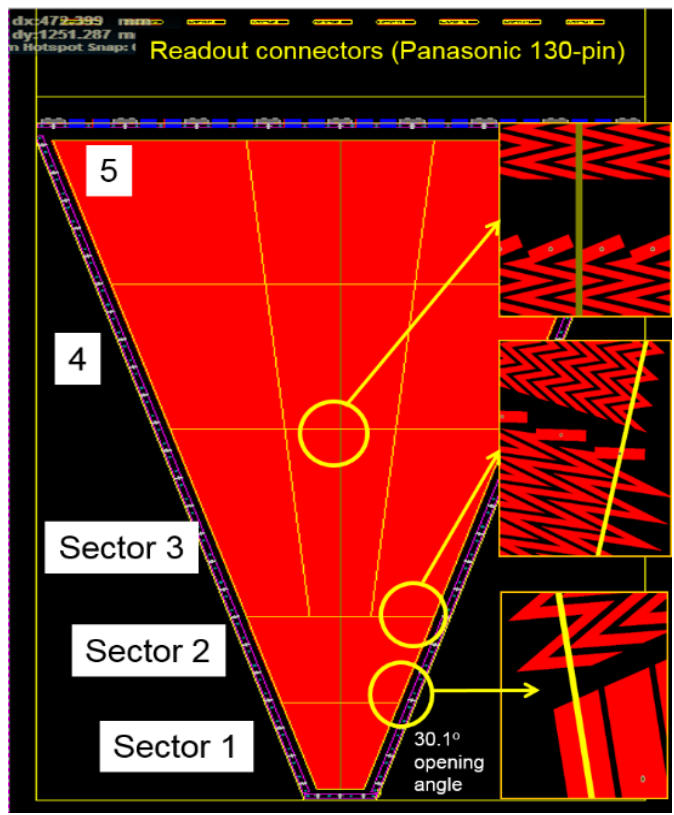

Fig. 6. Design of readout foil with radial zigzag strips.

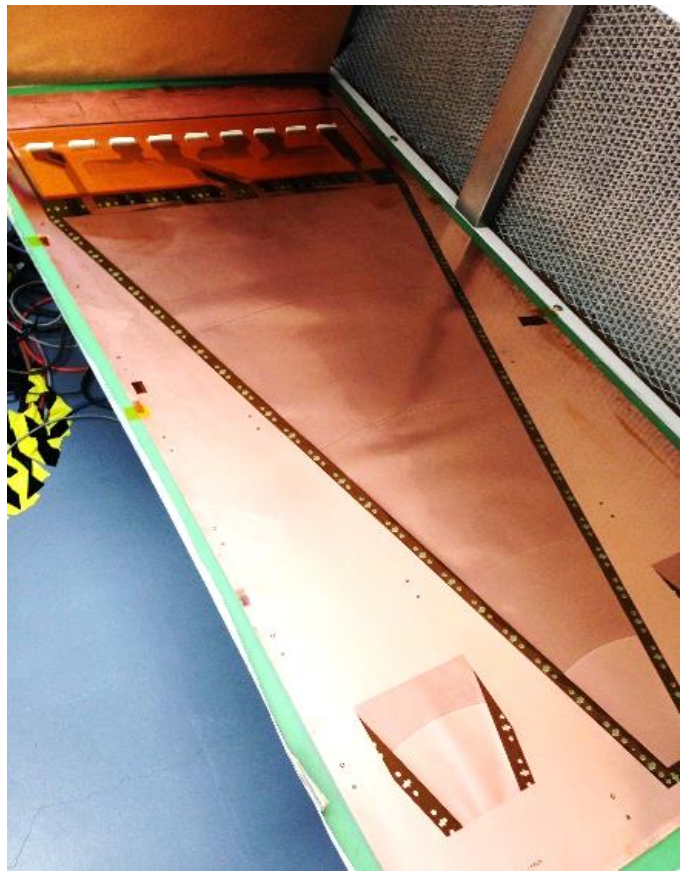

Fig. 7. Readout foil with radial zigzag strips produced at CERN.

## IV. SUMMARY AND OUTLOOK

We have designed a large low-mass Triple-GEM detector prototype for forward tracking at a future Electron-Ion Collider. All detector elements in the active area are implemented on foils. The total material in the active detector area is just below 0.6% and dominated by the copper cladding of the five foils. Light carbon fiber frames have been built to mechanically mount the foils and take up their tension. The readout foil features radial zigzag readout strips in four of the five radial sectors. We expect a spatial resolution of 50-90 µm at medium gas gains with this readout based on previous tests with small prototypes. The detector is currently under construction. We plan to commission and test the prototype with X-rays and particle beams in 2018.

ACKNOWLEDGMENT

We thank Prof. Ronnal P. Reichard, Dept. of Ocean Engineering and Science at Florida Institute of Technology, and the Structural Composites company for their assistance in the production of the carbon fiber frames.

We also thank Rui De Oliveira from the CERN PH-DT group for the production of GEM foils and readout foil.